\documentclass[aps,prb,twocolumn,superscriptaddress]{revtex4-2}

\usepackage{graphicx}
\usepackage{color}
\usepackage{braket}
\usepackage{amsfonts}
\usepackage{MnSymbol}
\usepackage{bm}
\usepackage{natbib}

\usepackage[dvipsnames]{xcolor}
\usepackage{mathrsfs}
\usepackage{mathtools}
\usepackage{subfigure}
\usepackage{comment}

\usepackage[colorlinks = true,linkcolor = blue,urlcolor  = blue,     citecolor = blue,anchorcolor = blue]{hyperref}

\usepackage{bbm}
\usepackage[dvipsnames]{xcolor}

\date{\today}                  

\begin{document}

\title{Current Flow in Topological Insulator Josephson Junctions due to Imperfections}

\author{Kiryl Piasotski}
\email[Email: ]{kiryl.piasotski@kit.edu}
\affiliation{Institute for Theory of Condensed Matter (TKM), 
Karlsruhe Institute of Technology,
76131 Karlsruhe, Germany}
\affiliation{Institute for Quantum Materials and Technologies (IQMT), 
Karlsruhe Institute of Technology, 76021 Karlsruhe, Germany}

\author{Omri Lesser}
\affiliation{Department of Condensed Matter Physics, Weizmann Institute of Science, 7610001 Rehovot, Israel}
\affiliation{Department of Physics, Cornell University, Ithaca, New York 14853, USA}

\author{Adrian Reich}
\affiliation{Institute for Theory of Condensed Matter (TKM), 
Karlsruhe Institute of Technology,
76131 Karlsruhe, Germany}

\author{Pavel Ostrovsky}
\affiliation{Max Planck Institute for Solid State Research, 70569 Stuttgart, Germany}

\author{Eytan Grosfeld}
\affiliation{Department of Physics, Ben-Gurion University of the Negev, Beer Sheva 84105, Israel}

\author{Yuriy Makhlin}
\affiliation{Condensed-matter physics Laboratory, HSE University, 101000 Moscow, Russia}
\affiliation{Landau Institute for Theoretical Physics, 142432 Chernogolovka, Russia}

\author{Yuval Oreg}
\affiliation{Department of Condensed Matter Physics, Weizmann Institute of Science, 7610001 Rehovot, Israel}

\author{Alexander Shnirman}
\affiliation{Institute for Theory of Condensed Matter (TKM), 
Karlsruhe Institute of Technology,
76131 Karlsruhe, Germany}
\affiliation{Institute for Quantum Materials and Technologies (IQMT), 
Karlsruhe Institute of Technology, 76021 Karlsruhe, Germany}

\begin{abstract}
Recent experiments on planar superconductor-topological insulator-superconductor (S-TI-S) junctions, e.g., in Corbino geometry, have reported low-temperature nonzero Josephson currents in states with integer fluxoid (flux) induced in the junction by a perpendicular magnetic field. 
This effect was discussed in connection with Majorana zero modes localized in Josephson vortices of such junctions. Here, we provide an explanation for this phenomenon, attributing it to imperfections. We focus on the ``atomic" limit in which the low-energy bound states of different vortices do not overlap. In this limit, we can associate the nonvanishing critical current with the irregularities, e.g., in the junction's width. The low-temperature contribution to the current is provided by the bound states with low but nonzero energy. We also propose clear experimental tests based on microwave spectroscopy, revealing distinctive selection rules for vortex transitions.

\end{abstract}

\maketitle

\section{Introduction}

The Majorana zero modes in artificial topological 
superconductors~\cite{kitaev2001unpaired}
have been in focus of both academic and industrial  
research for the last two decades. The leading 
proposals allowing for experimental realization are the 1-D semiconducting wires proximitized by regular superconductors  \cite{oreg2010helical,lutchyn2010majorana},
the extended planar (2-D) Josephson junctions 
on the surface of a 3-D topological insulator\cite{fu2008superconducting,fu2009josephson}(Fu \& Kane proposal) and the chains of magnetic adatoms on superconducting surfaces~\cite{Nadj-PerdePRB2013}. 
All these systems have by now been studied experimentally (see, e.g., Ref.~\onlinecite{aghaee2023inas} (and many earlier works) for 1-D wires,  Ref.~\onlinecite{yue2024signatures} for the planar Josephson junctions and Ref.~\onlinecite{Nadj-PerdeScience2014} for the chains of adatoms).

This paper is dedicated to the study of the planar (Fu \& Kane) platform and is motivated by the recent spark in the associated experimental activity~\cite{zhang2022ac,yue2024signatures, park2024corbino}. 
Whereas Ref.~\onlinecite{yue2024signatures}
addresses the Fraunhofer pattern in a long Josephson junction with open ends, in Refs.~\onlinecite{zhang2022ac,park2024corbino} long Josephson junctions of Corbino geometry were investigated. We focus here on the Corbino geometry as it allows to avoid the complications related to the boundary conditions at the open ends. Indeed, in the Josephson junctions with the open ends one measures~\cite{yue2024signatures} the standard Fraunhofer pattern since arbitrary magnetic flux is allowed in the junction. The simplest approach of integrating the local current-phase relation over the length of the junction might be a good first approximation, but might also miss the subtle effects related to the boundary conditions at the open ends, 
i.e., the 1-D Majorana modes having to hybridize there with the continuum of modes. The Corbino geometry avoids all these complications but, on the other hand, is subject to flux (fluxoid) quantization. Thus one cannot measure the full Fraunhofer pattern, but only the discrete points corresponding to the integer number of flux quanta, where naively the Josephson current should vanish.  

We aim at elucidating the recently observed~\cite{zhang2022ac, park2024corbino} non-vanishing Josephson currents in circular Corbino junctions when a non-zero number of flux quanta (fluxoids) are trapped in the ring. These should be closely related to the lifting of the zero nodes of the Fraunhofer pattern in the open-end junctions~\cite{yue2024signatures}.
The fact that these Josephson currents emerge only at temperatures much lower than the superconducting gap, induced in the TI, leads us to believe that they are related to the low-energy Andreev states in the Josephson vortices (a.k.a. Caroli-de Gennes-Matricon (CdGM) states\cite{CAROLI1964}).

To explain this phenomenon we examine the atomic limit of the Josephson vortex lattice created by an external magnetic field in the TI part of the junction~\cite{GrosfeldStern2011,potter2013anomalous,HegdeAnnPhys2020,BackensJETPLett2022}.
We observe that irregularities in, e.g., the width of the junction can explain the presence
of the observed Josephson currents, as conjectured in Ref.~\onlinecite{zhang2022ac}. Moreover, our simple 
estimates reproduce correctly the magnitude of the current observed in the experiment~\cite{park2024corbino}. 

We also study the current profiles associated with the individual CdGM states and show how these are modified by the imperfections. 
Finally, we investigate the microwave spectroscopy of the low-energy Andreev (CdGM) 
states. We predict very peculiar selection rules
for the allowed transitions, characteristic for the 
Josephson vortices in long topological Josephson junctions.  

\section{The system}
\begin{figure}[b!]
\centering
                \includegraphics[width=0.3\textwidth]{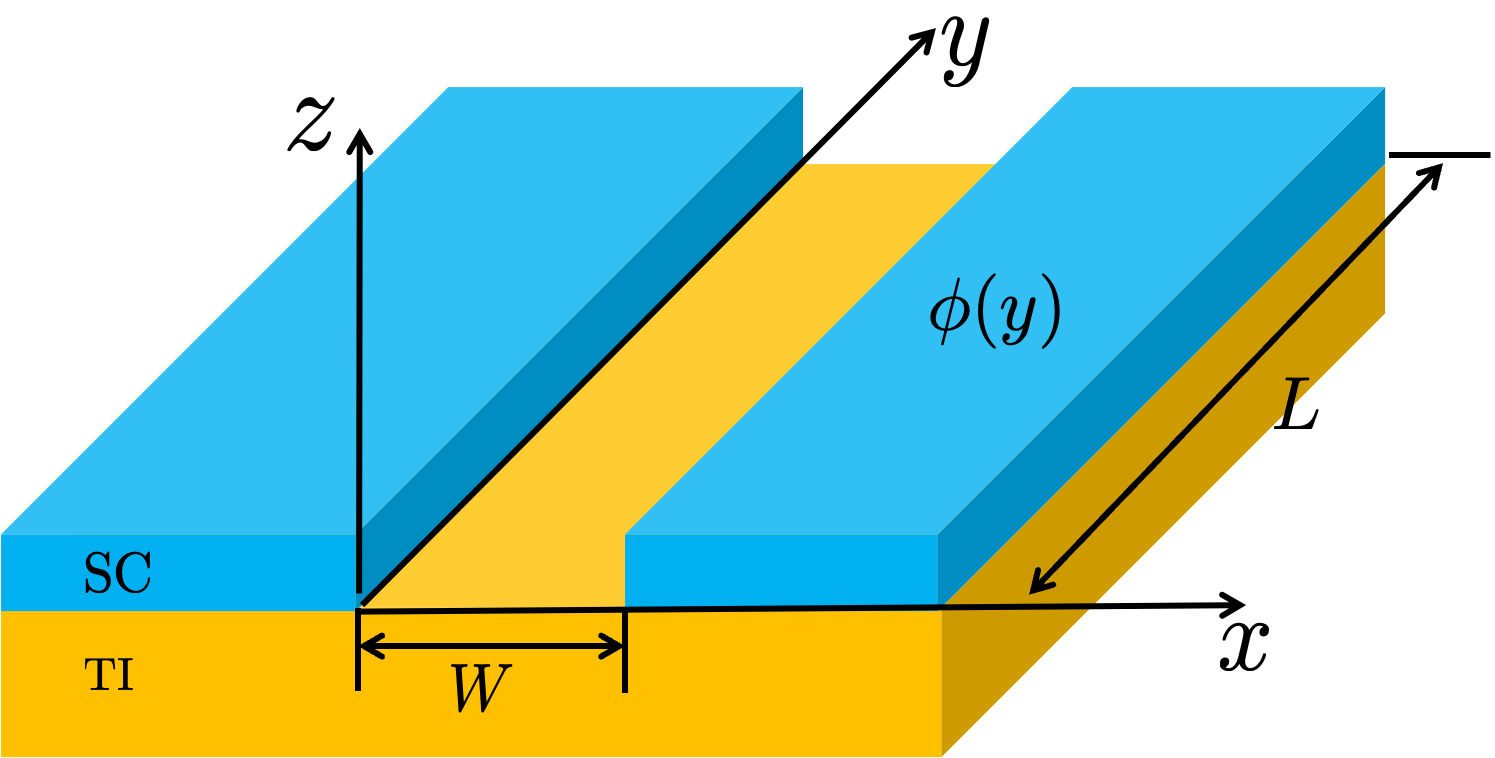}
                \includegraphics[width=0.15\textwidth]{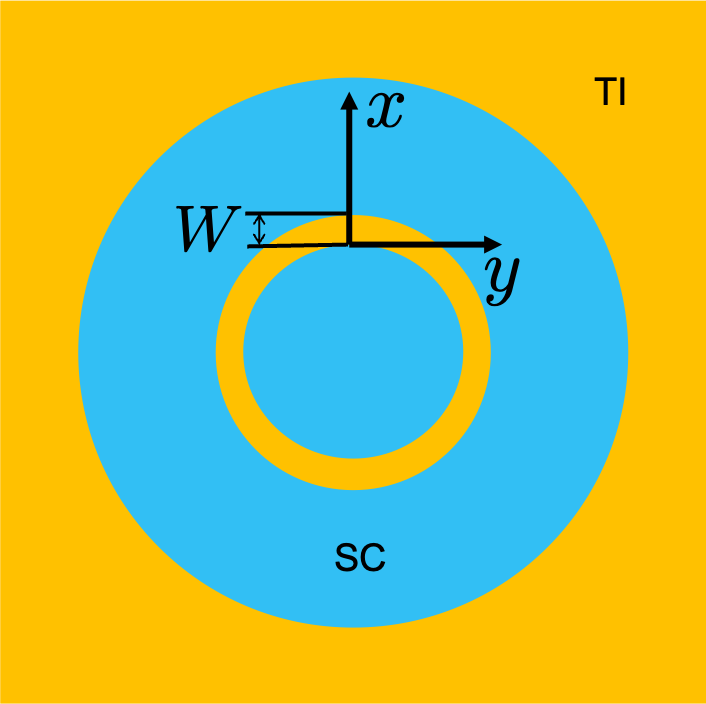}
                \caption{Left panel: Planar Josephson junction; right panel: Corbino geometry Josephson junction.}
                \label{fig: SystemL}
\end{figure}

We start by describing a planar Josephson junction shown in Fig.~\ref{fig: SystemL} (left panel).
Later we shall focus on the Corbino geometry and ``transform" this into a ring shape, see Fig.~\ref{fig: SystemL} (right panel).  
To model this system we consider the following Hamiltonian~\cite{fu2008superconducting,fu2009josephson,potter2013anomalous,hasan2010colloquium}
\begin{align}
    H=\frac{1}{2}\int{dx}\int_{0}^{L}{dy}\,\Psi^{\dagger}(\bold{r})h\Psi(\bold{r})\ , \label{microscopic_model}
\end{align}
where $\bold{r}\equiv (x,y)$,  $h=\tau_{z}\left(v{\bm\sigma}\cdot[\bold{p}+e\bold{A}(\bold{r})\tau_{z}]-\mu(\bold{r})\right)+(\Delta(\bold{r})\tau_{+}+\text{h.c.})$ is the corresponding Bogoliubov-de Gennes Hamiltonian, and $\Psi(\bold{r})=(\Psi_{\uparrow}(\bold{r}),\ \Psi_{\downarrow}(\bold{r}),\ \Psi_{\downarrow}^{\dagger}(\bold{r}),\ - \Psi_{\uparrow}^{\dagger}(\bold{r}))^T$ is the extended Nambu field. Here $v$ is the bare Fermi velocity associated with the topological insulator's Dirac cone, $\bold{p}=-i\boldsymbol{\nabla}=-i(\partial_{x},\ \partial_{y})$ is the momentum operator, ${\bm\sigma}=(\sigma_{x},\ \sigma_{y})$ is the vector of Pauli matrices operating on the spin space, while the $\tau$ matrices are the set of Pauli matrices associated with the particle-hole basis. 

\begin{figure}
\centering
                \includegraphics[scale=0.3]{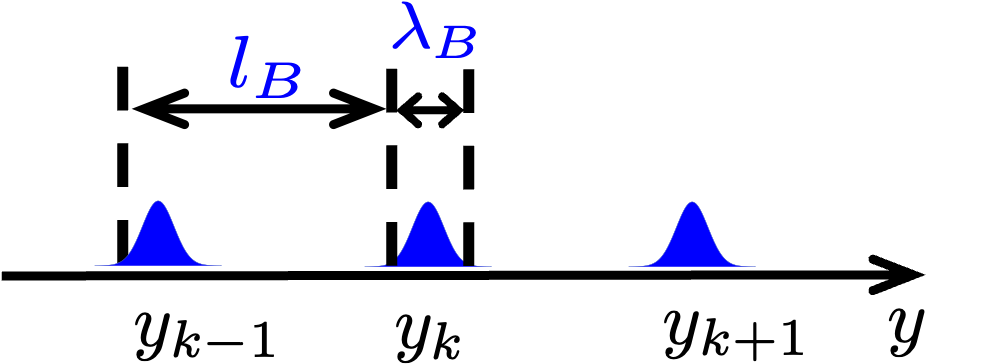}
                \caption{A cartoon showing a chain of vortices at distance $\ell_B$ from each other. The low-energy CdGM states are localized at distance $\lambda_B\ll \ell_B$.}
                \label{fig: AtomicChain}
\end{figure}

The proximity-induced superconducting energy gap $\Delta(\bold{r})$ is assumed to have a step-like profile $\Delta(\bold{r})=\Delta\Theta(-x)+e^{i\varphi(y)}\Delta\Theta(x-W)$, with the running phase $\varphi(y)=\frac{2\pi y}{\ell_{B}}$ arising from the external magnetic field treated in the Landau gauge~\footnote{Unlike in Ref.~\cite{GrosfeldStern2011} and similar to Refs.~\cite{potter2013anomalous,HegdeAnnPhys2020,BackensJETPLett2022} we consider only the case when the phase grows linearly with $y$. Thus, we assume the regime $\lambda_J\gg \ell_B$, where $\lambda_J$ is the Josephson penetration length~\cite{tinkham2004introduction}. This is justified if the effective Josephson energy of the junction is sufficiently small, or, alternatively, if the kinetic inductance of the superconducting films is sufficiently low.}. Specifically, irrespectively of the nature of the screening of the magnetic field by the super-current (London or Pearl regimes), we argue that the following choice is always possible: $A_{x}(\bold{r})=0$, $A_{y}(\bold{r})=A_{y}(x)$, and 
$\lim_{x\rightarrow -\infty}eA_{y}(x)=0, \quad \lim_{x\rightarrow\infty}eA_{y}(x)=-\frac{1}{2}\frac{d\varphi(y)}{dy}=-\frac{\pi}{\ell_{B}}
$, 
i.e., $A_{y}$ is there to compensate $\frac{d\varphi(y)}{dy}$ deep in the superconductor,
so that the super-current there vanishes (screening).
For example, under the assumption of London screening, we obtain $\frac{d\varphi(y)}{dy}=\frac{2\pi}{\ell_{B}}=\frac{2\pi}{\Phi_{0}}(2\lambda_{L}+W)B$, where $\lambda_{L}$ is the London's penetration depth.

To account for the renormalization of the chemical potential by the proximity to metallic superconductors, we follow Ref.~[\onlinecite{titov2006josephson}] and consider a step-like profile of the chemical potential $\mu(\bold{r})=\mu_{S}[\Theta(-x)+\Theta(x-W)]+\mu_{N}\Theta(x)\Theta(W-x)$, where, in the following, we assume $\mu_{S}$ to be the largest energy scale in our model (Andreev limit). 

The low-lying excitations of the topological Josephson junction described by the Hamiltonian of Eq. \eqref{microscopic_model} may be seen in certain regimes (to be specified below) as a quasi-one-dimensional lattice of Josephson vortex ``atoms''~\cite{GrosfeldStern2011,potter2013anomalous,HegdeAnnPhys2020,BackensJETPLett2022}. Each of these atoms is centered around $y\approx y_{k}= k\ell_{B}+\ell_{B}/2$, such that $\phi(y_{k})=(2k+1)\pi$. Each vortex hosts a number of localized orbitals (CdGM states). For the atomic limit to hold, the localization length of the low-energy orbitals $\lambda_B$ (to be defined later) must be much smaller than the distance between the vortices $\ell_B$, strongly suppressing the inter-vortex overlaps and, thus, rendering the vortex lattice into an array of nearly independent isolated atoms (see Fig.~\ref{fig: AtomicChain}). 

As we show in App.~\ref{App:EffectiveHamiltonian}, the $k^{\text{th}}$ atom is described by a two-component Nambu field $\psi_{k}(y)$ that is governed by the following effective $2\times 2$ Dirac (Bogoliubov-de Gennes) Hamiltonian:
\begin{align}
    h_{k}^{\text{eff}}=\frac{1}{2}\left\{-i\partial_y, v_{k}(y)\right\}\rho_{y}+\varepsilon_{k}(y)\rho_{z}. \label{hamiltonian}
\end{align}
In this Hamiltonian only the two lowest energy bound states of the transverse part of the microscopic Hamiltonian (see App.~\ref{App:EffectiveHamiltonian}) are taken into account. In the limit $W < \xi \equiv  v/\Delta$ considered in this paper these are the only 
in-gap states in the vicinity of the vortex center $y_k$.
The new set of Pauli matrices $\rho_x,\rho_y,\rho_z$ was introduced in order to avoid confusion with spin  $\vec \sigma$ and Nambu $\vec\tau$ Pauli matrices of the microscopic Hamiltonian~(\ref{microscopic_model}).
Here $\varepsilon_{k}(y)$ solves the following transcendental equation (see Ref.~[\onlinecite{PhysRevB.103.115423}]) 
$(\epsilon-i\sqrt{\Delta^{2}-\epsilon^{2}})^{2}=\Delta^{2} e^{-2i\epsilon W/v} e^{i\varphi(y)}$, 
for $y$ in the vicinity of $y_{k}$ (see Fig. \ref{fig: AtomicChain}). 

The $y$-dependent velocity $v_{k}(y)$, on the other hand, is deduced from the model parameters, including $\varepsilon_{k}(y)$, via the following formula (see App.~\ref{App:EffectiveHamiltonian})
\begin{widetext}
    \begin{align}
    v_{k}(y)=\frac{v}{1+\frac{W\tilde{\Delta}_{k}(y)}{v}}\left(\frac{\tilde{\Delta}_{k}(y)}{\mu_{N}}\sin\left(\frac{\mu_{N}W}{v}\right)+\frac{\tilde{\Delta}_{k}(y)}{\mu_{S}^{2}+\tilde{\Delta}^{2}_{k}(y)}\left[\tilde{\Delta}_{k}(y)\cos\left(\frac{\mu_{N}W}{v}\right)-\mu_{S}\sin\left(\frac{\mu_{N}W}{v}\right)\right]\right), \label{velocity_general}
\end{align}
\end{widetext}
where $\tilde{\Delta}_{k}(y)\equiv\sqrt{\Delta^{2}-\varepsilon_{k}^{2}(y)}$ determines the localization of the vortex states in the $x$-direction via $\sim e^{-\tilde{\Delta}_{k}(y)|x|/v}$. 

\begin{figure}
\centering
                
\includegraphics[scale=0.26]{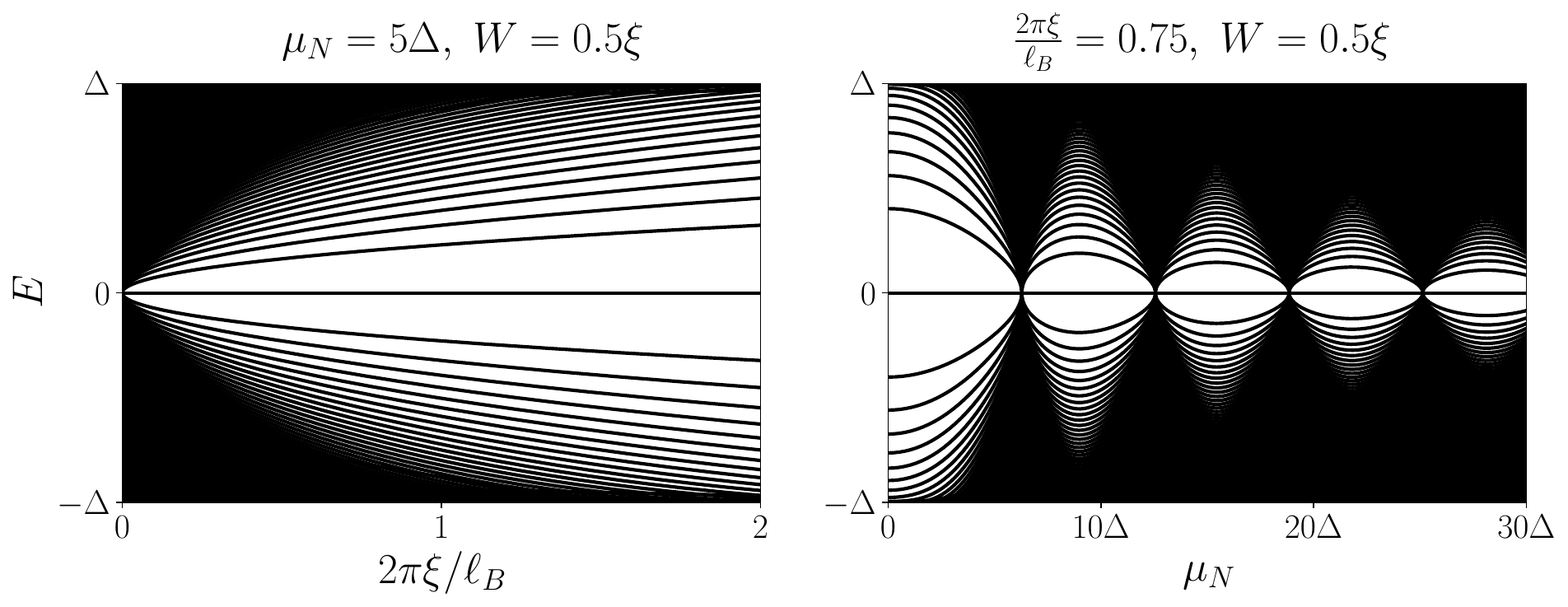}
                \caption{Spectral flow of a single topological Josephson vortex. The left panel shows the evolution of the spectrum with dimensionless magnetic field $\frac{2\pi \xi}{\ell_{B}}\propto B$. The right panel shows the evolution of the vortex spectrum with the chemical potential of the central region, $\mu_{N}$. Note that the spectrum condenses around the Fermi energy for the values of $\mu_{N}$ in resonance with the effective cavity modes $\mu_{N}=v\frac{\pi m}{W},\ m\in\mathbb{Z}$.}
                \label{fig: spectral_flow_bare}
\end{figure}

Before proceeding any further, we find it instructive to discuss Eq. \eqref{velocity_general} in greater detail. First, considering the Andreev limit $\mu_{S}\gg\max\{\Delta,\ \mu_{N},\ v/W\}$, we find 
\begin{align}
    v_{k}\simeq\frac{v}{1+\frac{W\tilde{\Delta}_{k}}{v}}\frac{\tilde{\Delta}_{k}}{\mu_{N}}\sin\left(\frac{\mu_{N}W}{v}\right), \label{Titov_Beenakker}
\end{align}
which at low energies $\tilde{\Delta}_{k}\simeq \Delta$ coincides with the result of Titov and Beenakker~\cite{titov2006josephson}. Note that in the limit $W\ll v/\mu_N$, Eq. \eqref{Titov_Beenakker} further simplifies to $v_{k}\simeq\tilde{\Delta}_{k}W$. 

Considering another limiting case of equal chemical potentials $\mu_{S}=\mu_{N}=\mu$, at low-energies $\tilde{\Delta}_{k}\simeq \Delta$, we recover
the result of Fu and Kane~\cite{fu2008superconducting}.
In this manner, Eq.~\eqref{velocity_general} serves to extend the previously established results. In the following sections, we will operate within the physically relevant Andreev limit described by Eq. \eqref{Titov_Beenakker}. 

Now we proceed to study the spectrum of the Hamiltonian \eqref{hamiltonian}. First, we note that this Hamiltonian possesses charge conjugation $C=\rho_{x}K$ and chiral $S=\rho_{x}$ symmetries. It has a topologically-protected zero-energy mode with the wave function given by
\begin{align}
    \chi_{k, 0}(y)\propto \begin{pmatrix}1 \\ \pm 1 \end{pmatrix}\frac{\exp\left(\mp\int^{y}dy'\frac{\varepsilon_{k}(y')}{v_{k}(y')}\right)}{\sqrt{v_{k}(y)}}\ ,
\end{align}
where the $\pm$ sign is chosen so that the resulting state is normalizable. 

To get an idea about the excited states (cf.~\cite{GrosfeldStern2011,potter2013anomalous,BackensJETPLett2022}), it is instructive to linearize $\epsilon_k(y)$ in the vicinity of $y=y_{k}$, i.e., $\epsilon_k(y) \approx\alpha_k\cdot (y-y_k)$ and fix the velocity as $v_k(y)\approx \bar v_k \equiv v_k(y_k)$. We obtain
\begin{align}\label{SuperOscillator}
h_{k}^{\text{eff}} \approx -i \bar v_k \rho_y\partial_y  +\alpha_k\cdot (y-y_k)\rho_z\ .
\end{align}
A simple estimate gives $\alpha_k \approx \frac{\Delta}{2}\frac{1}{1+W/\xi}\frac{2\pi}{\ell_B}$.
One observes~\cite{GrosfeldStern2011,potter2013anomalous,BackensJETPLett2022} that the resulting Hamiltonian 
is that of the super-symmetric oscillator, and may be diagonalized exactly 
(see App.~\ref{App:PerturbativeExpansion}). One finds that the excited states are exponentially localized in the Gaussian fashion $\propto\exp\left\{-\frac{(y-y_{k})^{2}}{2\lambda_{Bk}^{2}}\right\}$, with localization length given by
\begin{align}
    \lambda_{Bk}=\sqrt{\frac{\bar v_k}{\alpha_k}}\approx\Bigg{|}\frac{v\ell_{B}\sin(\mu_{N}W/v)}{\pi\mu_{N}}\Bigg{|}^{1/2}.
\end{align}
The RHS of this equation will depend on $k$ if $W$ or $\mu_N$ would become $y$-dependent as discussed below.
We observe that the atomic limit $\lambda_B\ll \ell_B$ requires either
$v/\mu_N \ll \ell_B$ and arbitrary $W$, or $W\ll \ell_B \ll v/\mu_N$.
For definiteness we will assume here the following experimentally relevant hierarchy of lengths:
$W\ll \ell_B\sim \xi \ll v/\mu_N$. Then, $\bar v_{k} \sim \Delta W \sim v W/\xi \ll v$. Recall we assume the 
Andreev limit $\mu_{S}\gg\max\{\Delta,\ \mu_{N},\ v/W\}$ leading to Eq.~(\ref{Titov_Beenakker}).

For the energy levels, one finds the following square root scaling $E_{n,k}=-E_{-n,k}=\omega_{Bk}\sqrt{n}, \ n=1, 2, 3\dots$, with the frequency given by
\begin{align}
\omega_{Bk}=\sqrt{2\bar v_k\alpha_k}=\frac{\Delta}{1+W/\xi}\Bigg{|}\frac{2\pi v\sin(\mu_{N}W/v)}{\mu_{N}\ell_{B}}\Bigg{|}^{1/2}. \label{bare frequency}
\end{align}

Fig.~\ref{fig: spectral_flow_bare} shows the spectrum of a single Josephson vortex evaluated numerically by diagonalizing the Hamiltonian (\ref{hamiltonian}). In particular, panel (a) demonstrates the spectral flow of Andreev levels with the dimensionless magnetic field $\frac{2\pi\xi}{\ell_{B}}$, while panel (b) exemplifies that with the chemical potential in the junction region $\mu_{N}$. As is apparent from Fig.~\ref{fig: spectral_flow_bare}, our low-energy formula \eqref{bare frequency} carefully captures both $\frac{2\pi\xi}{\ell_{B}}$ and $\mu_{N}$ dependencies of the spectrum.

We observe that irregularity could make 
the width $W(y)$ or the chemical potential $\mu_N(y)$ dependent on $y$. 
Then the eigenfrequencies $\omega_{Bk}$ and the localization 
lengths $\lambda_{Bk}$ would indeed depend on the location 
$y_k$ of each vortex (cf. pinning) and would vary between vortices.

\section{Josephson current due to variation of the width}
We close the system into a ring of length $L=N\ell_{B}$ in $y$-direction.
In the experiment, $N$ and $\ell_B = L/N$ are determined by the external magnetic field, screening of the magnetic field (London or Pearl regimes), and the fluxoid quantization. The boundary conditions emerging upon encircling the ring are not important in the atomic limit~\footnote{These are very important when the CdGM states of the vortices overlap.}.
This way we model the Corbino disk geometry junctions recently studied experimentally in Refs.~\onlinecite{zhang2022ac, park2024corbino}. 

We start by pointing out that without irregularities the total current in the system hosting an integer number $N=\frac{L}{\ell_{B}}\neq 0$ of flux quanta vanishes exactly, as the externally induced phase difference $\varphi_{0}$, entering the problem via $\varphi(y)=\frac{2\pi y}{\ell_{B}}+\varphi_{0}$, may be completely removed from consideration by the coordinate transformation $y\rightarrow y-\frac{\varphi_{0}}{2\pi}\ell_{B}$ (remember we are on a ring), thus rendering the Gibbs free energy independent of $\varphi_{0}$ and leading to the null Josephson current $J=\frac{2\pi }{\Phi_{0}}\partial_{\varphi_{0}}F=0$. In contrast, if we allow any of the model parameters to have additional $y$-dependence or introduce any $y$-dependent perturbation, we cannot eliminate the external offset phase $\varphi_{0}$, which inevitably leads to the emergence of non-zero critical currents.

Perhaps, one of the simplest things to imagine is to assume that the width $W$ admits for small variations along the junction $W=W(y)$. 
Another possibility are random gate charges leading to variations of $\mu_N(y)$.

Assuming such a variation to be slow on the scale of $\lambda_{Bk}$, we may roughly assume the frequency of the $k^{\text{th}}$ oscillator to change as
\begin{align}
    \omega_{B, k}(\varphi_{0})\approx \frac{\Delta}{1+W_{k}(\varphi_{0})/\xi}\Bigg{|}\frac{2\pi v\sin(\mu_{N}W_{k}(\varphi_{0})/v)}{\mu_{N}\ell_{B}}\Bigg{|}^{1/2},
\end{align}
where $W_{k}(\varphi_{0})=W\left(y_{k}+\frac{\varphi_{0}}{2\pi}\ell_{B}\right)$, giving rise to a non-zero current
\begin{align}
    I(\varphi_{0})=-\frac{\pi}{\Phi_{0}}\sum_{k}\sum_{n=1}^{n_{\text{max}}}\tanh\left(\frac{\beta E_{n, k}}{2}\right)\frac{\partial E_{n, k}}{\partial \varphi_{0}}. \label{current_disorder}
\end{align}
In principle, $n_{\text{max}}$ could be the number of energy 
levels that fit into the gap. This is a subtle issue as the continuum states over the gap could also contribute to the current. Here we focus on the contribution of the low energy states only, $E_{n,k} \ll \Delta$, as these can be extracted by performing measurement at different temperatures.

Indeed, this simple consideration allows us to make a good connection with a recent experiment~\cite{park2024corbino}. In that work, the critical current $I_{c}=\max_{\varphi_{0}\in[0, 2\pi)}I(\varphi_{0})$ is measured at various temperatures. The experiment reveals that a drop in the temperature from $T=1.6\,\rm{K}$ to $T=0.25\,\rm{K}$ produces an increase in the $N\neq 0$ critical current from being roughly zero $I_{c}\sim 0$ to around $I_{c}\sim 10\,\rm{nA}$.

This result, when analyzed with the help of Eq.~\eqref{current_disorder}, immediately tells us that the observed effect was produced by the low-lying excitations of the system, as it is their contribution that gets enhanced by the $\tanh\left(\frac{\beta E}{2}\right)$ function upon the temperature drop. Using the experimentally observed value of the zero-field ($N=0$) critical current $I_{c}(B=0)\sim 2\,\rm{\mu A}$ (see Ref.~\onlinecite{park2024corbino}), as well as a rough estimate of the Corbino disk dimensions $L/W\sim20$, we may recover the size of the energy gap from the approximate relation~\cite{titov2006josephson} $I_{c}(B=0)\sim \frac{\pi\Delta}{\Phi_{0}}\frac{L}{W}$ to be $\frac{\pi\Delta}{\Phi_{0}}\sim 100\,$nA ($\Delta/k_B\approx 5$K). For the contribution of the first CdGM state to the Josephson current at $T\ll E_{1,k} = \omega_{Bk}$ we obtain
\begin{align}
   \delta I_{1} = -\frac{\pi}{\Phi_{0}}\frac{\partial E_{1,k}}{\partial \varphi_{0}}&\sim -\frac{\pi\Delta}{2\Phi_{0}}\sqrt{\frac{\ell_{B}}{2\pi W}}\frac{\partial W}{\partial y_{k}}\ .
\end{align}
Introducing a typical variation of the width 
$\delta W$ and estimating roughly $|\partial W/\partial y_{k}| \sim \delta W/L = \delta W/(N \ell_B)$ we obtain
\begin{align}
|\delta I_{1}|
    &\sim \frac{\pi\Delta}{\Phi_{0}}\sqrt{\frac{W}{8\pi N L}}\frac{\delta W}{W}\sim 0.5\,{\rm nA}/\sqrt{N}\ . \label{estimate}
\end{align}
We have allowed for the width variations of the order of $10\%$, that is $\delta{W}/W\sim0.1$. 
Taking into account that several low-energy
CdGM states can contribute and since $E_{n,k} = \sqrt{n} E_{1,k}$, i.e., $\delta I_n \sim \sqrt{n} \,\delta I_1$, we see that our simple logic gives rise to a correct quantitative estimate of the observed current~\cite{park2024corbino}. An example of width irregularity is provided in App.~\ref{App:ExampleIrregularity}.

\section{Current profiles of individual CdGM states}
To elucidate how irregularities of the width $W$ or the chemical 
potential $\mu_N$ generate Josephson current, we investigate here the 
contributions of the individual CdGM levels. We start 
again with the Hamiltonian of a single vortex (\ref{hamiltonian}). 
To derive the current density operator, one 
can formally subject the system to an external flux $\varphi(y)\rightarrow \varphi(y)+\frac{2\pi\Phi(t)}{\Phi_{0}}$ (see Refs.~\onlinecite{Devoret2020, Devoret2021, Houzet2024}), which, to the lowest order in perturbation theory, modifies the effective Hamiltonian \eqref{hamiltonian} as
\begin{align}
    h_{k}^{\text{eff}}(t)=-\frac{1}{2}\left\{v_{k}(y), p_{y}\right\}\rho_{y}+\varepsilon_{k}(y)\rho_{z}-j_{k}(y)\Phi(t)\ ,\label{hamiltonian_drive}
\end{align}
where $j_{k}(y)$ is the projection of the current density operator onto the low-energy subspace of the $k^{\text{th}}$ vortex and is given by (see App.~\ref{App:CurrentOperator})
\begin{align}
    j_{k}(y)=\frac{\ell_{B}}{\Phi_{0}}\frac{\partial \varepsilon_{k}(y)}{\partial y}\rho_z\ . \label{current_vertex}
\end{align}

Next, motivated by the approximate relation $v_k(y) \sim \Delta W(y)$, we linearize around $y=y_k$
and obtain $v_k(y) \approx \bar v_k + \gamma_k\cdot (y-y_k)$, where $\gamma_k = \Delta \partial W/\partial y|_{y=y_k}$. Assuming, as in (\ref{SuperOscillator}), $\epsilon_k(y) = \alpha_k\cdot(y-y_k)$ we obtain 
\begin{align}
j_{k}(y) = (\ell_B\alpha_k/\Phi_0)\rho_z\ ,
\end{align}
and
\begin{align}\label{eq:hkeff}
\nonumber
    h_{k}^{\text{eff}}&=\frac{1}{2}\left\{-i\partial_y, v_{k}(y)\right\}\rho_{y}+\varepsilon_{k}(y)\rho_{z} \\
    &\approx -i \bar v_k \rho_y\partial_y  +\alpha_k\cdot (y-y_k)\rho_z+\frac{\gamma_k}{2}\left\{-i\partial_y,(y-y_k)\right\}\rho_{y}\ .
\end{align}

\begin{figure}[t!]
\centering
 \includegraphics[width=0.35\textwidth]{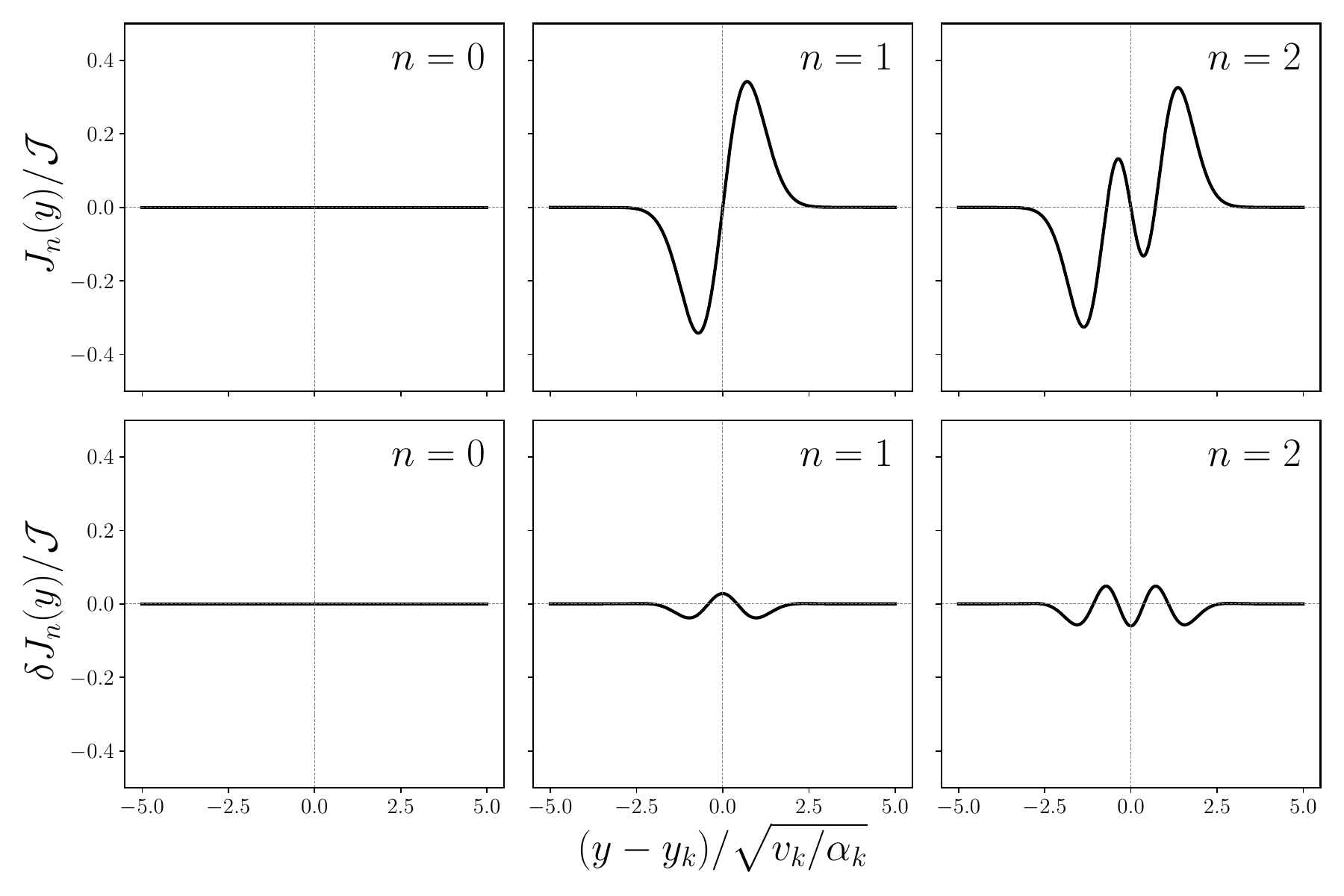}
                \caption{Current profiles of the individual CdGM states. The upper panels correspond to the unperturbed states. The lower panels show the corrections (with $\gamma_k/\omega_{Bk}=0.05$) due to the gradient of the width $W(y)$.}
                \label{fig: current profiles}
\end{figure}

We analyze (\ref{eq:hkeff}) perturbatively (see App.~\ref{App:PerturbativeExpansion}) and obtain two important conclusions: a) The current profile of the zero (Majorana) level 
does not get any correction and remains zero; 
b) The current profile of the first level, as well as of all higher levels, get corrected. Moreover, the corrections 
contain contributions that produce a finite result if integrated over $y$, $I_n = \int dy \,\delta J_n(y) = I_1 \sqrt{n}$, where $I_1 \equiv -\mathcal{J}\gamma_k/\omega_{Bk}$.
The results are shown in Fig.~\ref{fig: current profiles}.

\section{Andreev spectroscopy}
The predicted spectral properties of the junction may be experimentally assessed via microwave spectroscopy techniques~\cite{bretheau2013supercurrent, tosi2019spin, Kurilovich2021, bargerbos2023spectroscopy, wesdorp2024microwave}. To drive the transitions between the Andreev levels~\footnote{An alternative to the flux (current) drive (see Refs.~\cite{Devoret2020, Devoret2021, Houzet2024}, for example) would be the gate drive as in Refs.~\cite{bargerbos2023spectroscopy, Nazarov2010}}, one subjects the system to an external time-dependent flux (now not only formally) which leads again to Eqs.~(\ref{hamiltonian_drive}) and (\ref{current_vertex}).

It follows that to understand the transitions in our system we have to analyze the matrix elements of the current operator in Eq. \eqref{current_vertex} because it multiplies the externally applied time dependent flux in Eq.~\eqref{hamiltonian_drive}. We thus evaluate
$
    I_{n, n'}=\sum_{k}\int_{0}^{L}{dy}\bra{\Psi_{n,k}(y)}j_{k}(y)\ket{\Psi_{n',k}(y)}
$.

Naturally, we find that $I_{n, n}=0$, that is there is no net current without disorder, as discussed above. 
We find that the only non-zero matrix elements are 
\begin{align}
    I_{n, n+1}=&I_{n, -(n+1)}=-I_{n+1, -n}=-I_{-(n+1), -n}\approx \frac{1}{2}\tilde{I},\nonumber \\
    I_{0, 1}=&I_{0, -1}=-I_{-1, 0}=-I_{1, 0}\approx\frac{1}{i\sqrt{2}}\tilde{I},
\end{align}
where $\tilde{I}=\frac{\Delta}{1+W/\xi}\frac{\pi}{\Phi_{0}}$.

\begin{figure}[t!]
\centering
 \includegraphics[scale=0.32]{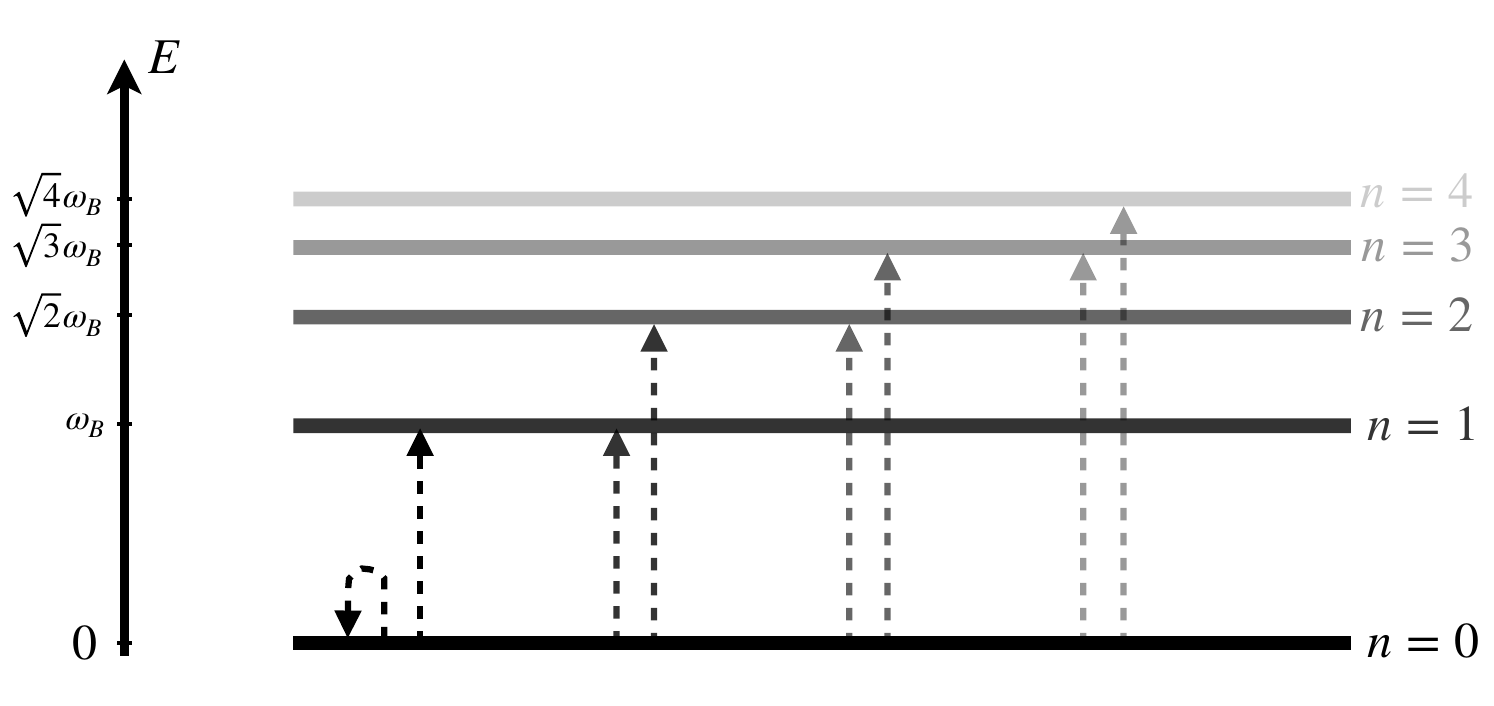}
                \caption{The schematic depiction of the allowed transitions in vortex atoms. In particular, by driving the system at frequencies $\Omega=\omega_{B}(\sqrt{n}+\sqrt{n-1}),\ n=1, 2, \dots$ it is possible to break Cooper pairs in the ground state to populate pairs of neighboring excited states at energies $\omega_{B}\sqrt{n}$ and $\omega_{B}\sqrt{n-1}$.}
                \label{fig: transitions}
\end{figure}

This analysis indicates that at absolute zero, absorption transitions happen at frequencies $\Omega_{n,k}=\omega_{Bk}(\sqrt{n}+\sqrt{n-1}),\ n=1, 2, \dots$, involving the pairwise population of neighboring CdGM levels with energies $\omega_{Bk}\sqrt{n}$ and $\omega_{Bk}\sqrt{n-1}$, by two quasiparticles (see Fig. \ref{fig: transitions}). For $n=1$ this transition requires flipping the parity of the Majorana zero energy  state (more precisely of the fermion state formed by this zero-energy level and by another one, which could be at a different vortex or at the system's boundary).

\section{Conclusions and outlook}

In this manuscript, we present a study of the properties of Caroli-de Gennes-Matricon (CdGM) states associated with Josephson vortex atoms, where the CdGM states of different vortices remain non-overlapping, in topological Josephson junctions. Our analysis provides insights into the recently observed non-zero critical currents in junctions hosting an integer number of vortices, linking this phenomenon to the response of low-lying CdGM states to perturbations that disrupt translational symmetry at the scale of a single magnetic length. We have investigated the current profiles of individual CdGM states under such perturbations and found that the current profiles of all states, except the zero-energy Majorana state, are modified, resulting in a finite total current. Additionally, we propose a method to probe the low-lying excitations of topological Josephson junctions through microwave spectroscopy, predicting a particularly clear pattern of resonances in the system.  

Notice that in this paper, we do not discuss the geometric (Shapiro-like) effects leading to the appearance of relatively large Josephson currents (at relatively high temperatures) in square-shaped Corbino junctions when the number of flux quanta is a multiple of four~\cite{park2024corbino}. Here, we concentrate on a weaker geometric effect in strictly circular Corbino junctions, i.e., the irregularity of the junction's width. This leads to much smaller Josephson currents, which show up at very low temperatures at arbitrary numbers of flux quanta~\cite{park2024corbino}.

Several questions and intriguing directions were not addressed in the current research and should be explored in future studies.
How do the results change beyond the atomic limit when there is an overlap between the CdGM states? What is the impact of Zeeman coupling, and what role do fluctuations in the phase of the order parameter play? Can these fluctuations mediate interactions between the CdGM states that are stronger than the exponentially small hybridization?
We anticipate that in the long junction limit, it may be crucial to incorporate phase fluctuations in a self-consistent manner, considering the back-reaction of the current on the magnetic field. It would also be interesting to 
study the effect of the inhomogeneities of the chemical potential $\mu_{N}$ and of the induced gap $\Delta$ on the Josephson current.

In the last stages of preparation for this manuscript, a closely related manuscript was submitted~\cite{Laubscher2024}. In contrast to the study of Ref.~\onlinecite{Laubscher2024} of open-ended geometry, here we discuss the Corbino geometry. We arrive at similar results for the microwave spectroscopy.
The authors of Ref.~\onlinecite{Laubscher2024} indicate that the lifting of the zeros in the Fraunhofer pattern may be related to the boundary conditions at the open ends. Our results in the current manuscript suggest that irregularities could also play a role in lifting these zeros.

\section*{Acknowledgments}

This paper has greatly benefited from the discussions with Stefan Backens.
AS acknowledges useful discussions with Viktor Yakovenko. We acknowledge funding from the DFG Project SH 81/7-1. Y.O. acknowledges support by Deutsche Forschungsgemeinschaft through CRC 183 and by the ISF 2478/24 grant.

\bibliography{citations}

\appendix
\begin{widetext}

\section{Effective single-vortex Hamiltonian}
\label{App:EffectiveHamiltonian}

\begin{figure}[h]
\centering
                \includegraphics[scale=0.4]{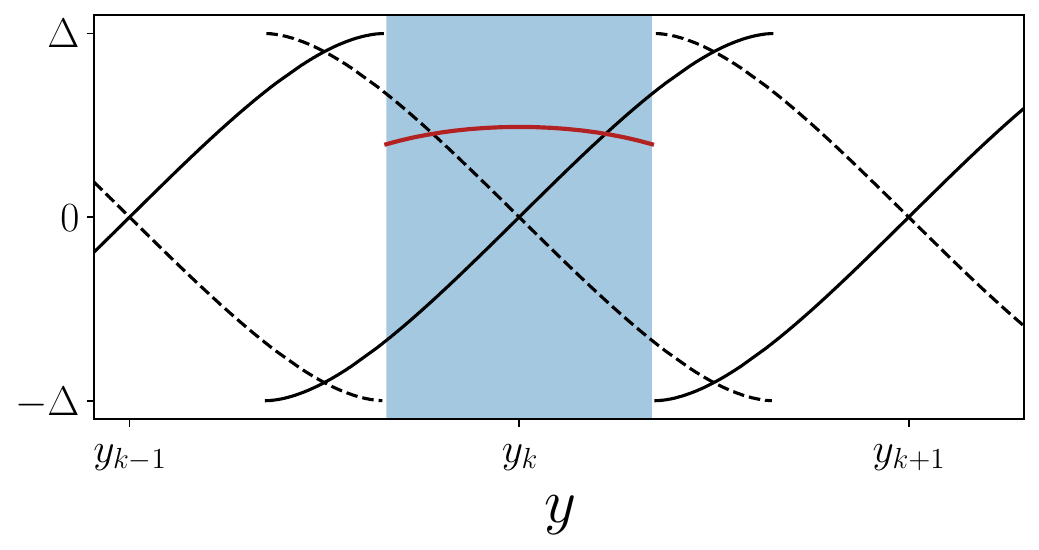}
                \caption{The solid ($\sigma=+$) and the dashed ($\sigma=-$) black lines are the eigenenergies of the  auxiliary Hamiltonian of Eq. \eqref{LET_unperturbed}. The blue area indicates roughly the domain of one vortex. The red line shows a typical profile of the velocity $v_k(y)$.}
                \label{specs}
\end{figure}
To derive an effective Hamiltonian of a single vortex, we follow Ref.~\cite{BackensJETPLett2022} and consider first the eigenstates of 
\begin{align}
    h_{0}=v\sigma_{x}\tau_{z}p_{x}-\mu(x)\tau_{z}+\left(\Delta(x, y)\tau_{+}+\text{h.c.}\right), \label{LET_unperturbed}
\end{align}
treating the kinematics in $y$-direction $h_{1}=v\tau_{z}\sigma_{y}[p_{y}+eA_{y}(x)\tau_{z}]$ as a perturbation.

Assuming $W\ll\xi$, in the vicinity of $y\approx y_{k}=\ell_{B}/2+k\ell_{B}$ (the blue region in Fig. \ref{specs}), we approximate the wave function as a superposition of the ($\sigma=+$) and ($\sigma=-$) states 
\begin{align}
    \Psi(x, y)\approx \sum_{\sigma=\pm}\alpha_{\sigma, k}(y)\psi_{\sigma, k}^{(0)}(x, y),\quad y\simeq y_{k}, \label{expansion}
\end{align}
where $\psi_{\sigma,k}^{(0)}(x, y)$ are the eigenfunctions of $h_{0}$ defined as

\begin{align}
    \nonumber
    \psi_{\sigma, k}^{(0)}(x, y)=&\sqrt{\frac{\kappa_{k}(y)}{2[1+\kappa_{k}(y)W]}}e^{i\frac{\pi y}{2\ell_{B}}\tau_{z}}\Bigg[\Theta(-x)e^{i\sigma \mu_{S} x/v}e^{-i\sigma\mu_{N}W/2v}\begin{pmatrix}e^{-i\epsilon_{k}(y) W/2v}  \\ \sigma  e^{i\epsilon_{k}(y) W/2v}  \end{pmatrix}e^{\kappa_{k}(y) x}\\
    \nonumber
    &+\Theta(x)\Theta(W-x)e^{i\sigma \mu_{N} (x-W/2)/v}\begin{pmatrix}e^{i\epsilon_{k}(y) (x-W/2)/v} \\ \sigma e^{-i\epsilon_{k}(y) (x-W/2)/v}  \end{pmatrix} \\ 
    &+\Theta(x-W)e^{i\sigma \mu_{S} (x-W)/v}e^{i\sigma\mu_{N}W/2v}\begin{pmatrix} e^{i\epsilon_{k}(y) W/2v} \\ \sigma  e^{-i\epsilon_{k}(y) W/2v}\end{pmatrix}e^{-\kappa_{k}(y) (x-W)}\Bigg]\ket{\sigma_{x}=\sigma},
\end{align}    
where $\sigma\varepsilon_{k}(y)$ is the corresponding eigenenergy shown in Fig. \ref{specs}. The $y$-dependent localization parameter $\kappa_{k}$, is defined through
\begin{align}
    v\kappa_{k}(y)=\sqrt{\Delta^{2}-\varepsilon^{2}_{k}(y)}\equiv \Delta_{k}(y).
\end{align}

Substituting the ansatz \eqref{expansion} into the Schr\"odinger equation $(h_{0}+h_{1})\Psi=E\Psi$, we discover 
\begin{align}
    \sum_{\sigma'=\pm}[v_{k}^{\sigma, \sigma'}(y)p_{y}\alpha_{\sigma', k}(y)+a_{k}^{\sigma, \sigma'}(y)\alpha_{\sigma', k}(y)]+\sigma\varepsilon_{k}(y)\alpha_{\sigma, k}(y)=E\alpha_{\sigma, k}(y),
\end{align}
where
\begin{align}
\nonumber
   v_{k}^{\sigma, \sigma'}(y)=&\int{dx}\psi^{\dagger}_{\sigma}(x, y-y_{k})v\tau_{z}\sigma_{y}\psi^{\phantom\dagger}_{\sigma'}(x, y-y_{k})=\\
   =&\frac{-i\sigma\delta_{\sigma', -\sigma}v}{1+\frac{W(y)\tilde{\Delta}_{k}(y)}{v}}\left(\frac{\tilde{\Delta}_{k}(y)}{\mu_{N}}\sin\left(\frac{\mu_{N}W(y)}{v}\right)+\frac{\tilde{\Delta}_{k}(y)}{\mu_{S}^{2}+\tilde{\Delta}^{2}_{k}(y)}\left[\tilde{\Delta}_{k}(y)\cos\left(\frac{\mu_{N}W(y)}{v}\right)-\mu_{S}\sin\left(\frac{\mu_{N}W(y)}{v}\right)\right]\right),\\
   a_{k}^{\sigma, \sigma'}(y)=&\int{dx}\psi^{\dagger}_{\sigma}(x, y-y_{k})h_{1}\psi^{\phantom\dagger}_{\sigma'}(x, y-y_{k})=\frac{1}{2i}\frac{\partial v_{k}^{\sigma, \sigma'}(y)}{\partial y}.
\end{align}
In this respect, the sought differential equation for the expansion coefficients $\alpha_{\sigma, k}$ becomes
\begin{align}
   \left(-\frac{1}{2}\{v_{k}(y), p_{y}\}\rho_{y}+\varepsilon_{k}(y)\rho_{z}\right)\begin{pmatrix}\alpha_{+, k}(y)\\  \alpha_{-, k}(y)\end{pmatrix}=E\begin{pmatrix}\alpha_{+, k}(y)\\  \alpha_{-, k}(y)\end{pmatrix}. \label{Dirac_app}
\end{align}

\section{Example of irregularity}
\label{App:ExampleIrregularity}

In Figs.~\ref{fig: 1impurity} and \ref{fig: 1impurity_currents}, we demonstrate a particular realization of an irregular $W(y)$.

\begin{figure}[t!]
\centering
\includegraphics[scale=0.15]{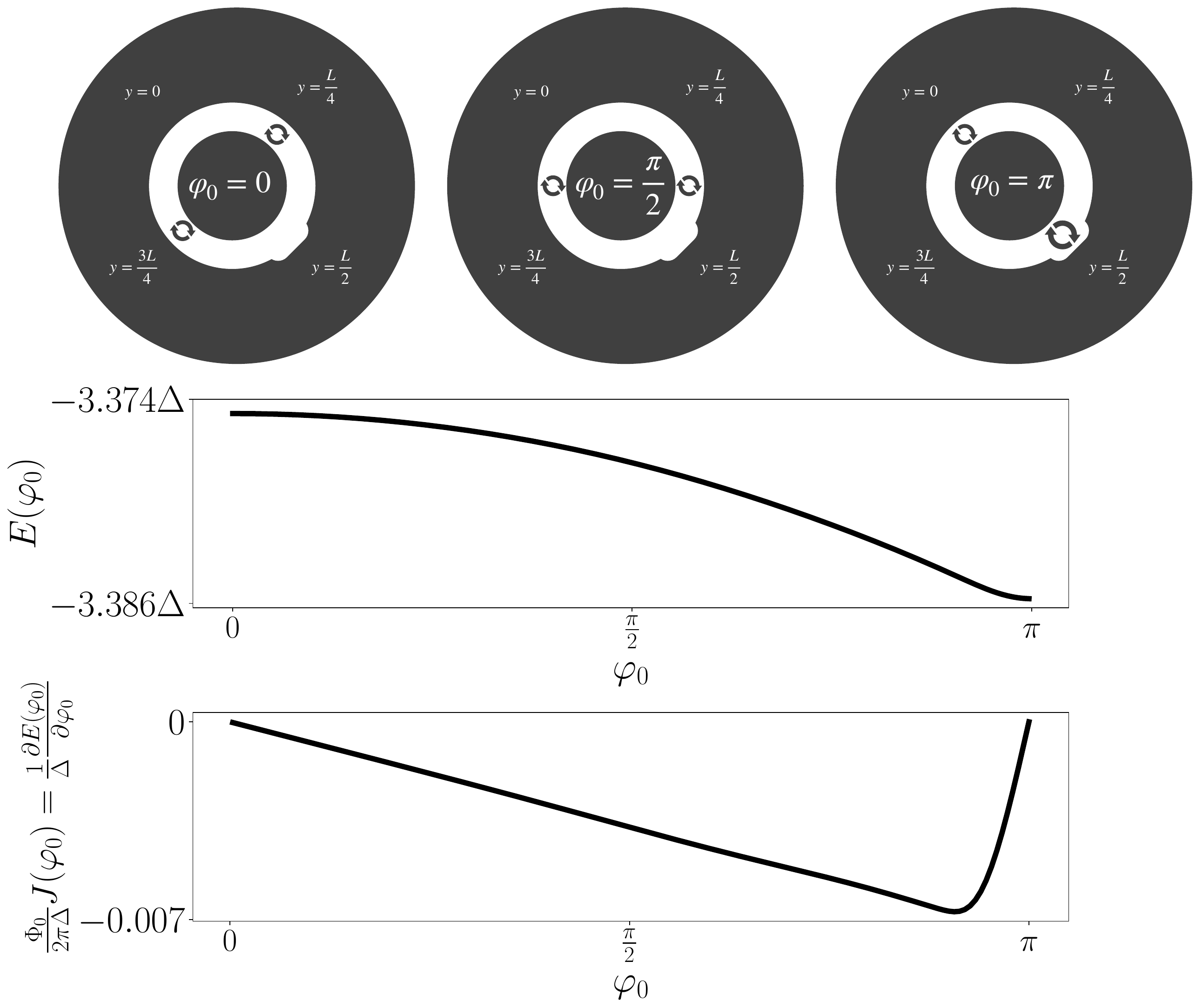}
\caption{This figure illustrates how the energy of Josephson vortices in a Corbino junction with a single extended impurity (widening) varies according to their position, set by the external phase difference $\varphi_{0}$. 
The upper panel is a cartoon of a Corbino junction with a single extended widening and 
two trapped vortices.
The middle panel shows the contribution of the CdGM states to the ground state energy $E(\varphi_{0})=-\frac{1}{2}\sum_{n=1}^{n_{\text{max}}}E_{n}(\varphi_{0})$, while the lower panel shows the corresponding Josephson current $J(\varphi_{0})=\frac{2\pi}{\Phi_{0}}\frac{\partial E(\varphi_{0})}{\partial \varphi_{0}}$. As one would naively expect, a local increase in the junction width $W$, locally increases the oscillator frequencies $\omega_{Bk}/\Delta\sim \sqrt{W/\ell_B}$, reducing the total ground state energy of the system. The simulations were performed in the regime $W\ll\xi$ with $L=40 W$ and a single widening located at $y=L/2$, spread over the length of $\frac{3}{10\pi}L$, and with $\delta{W}=W/10$. The order of magnitude of the Josephson current is consistent with the estimate (\ref{estimate}).}
\label{fig: 1impurity}
\end{figure}

\begin{figure}[t!]
\centering
\includegraphics[scale=0.15]{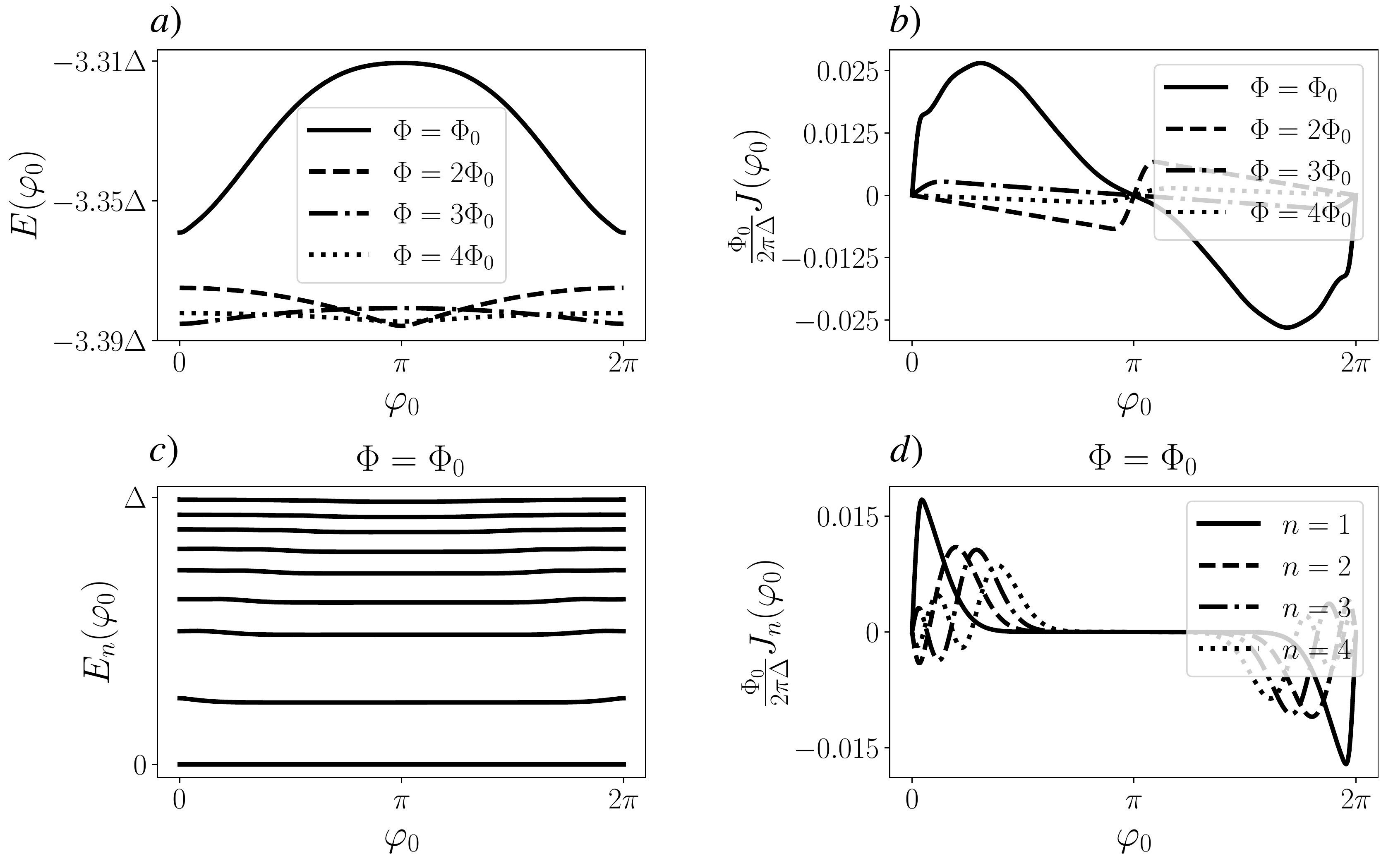}
\caption{Data for the Corbino junction with a single extended impurity (see Fig.~\ref{fig: 1impurity}). Panel a): The contribution of the CdGM states to the ground state energy $E(\varphi_{0})=-\frac{1}{2}\sum_{n=1}^{n_{\text{max}}}E_{n}(\varphi_{0})$ for the total flux (fluxoid) $\Phi=N\Phi_{0}$, where 
$N=1,2,3,4$ is the number of trapped vortices. Panel b): The Josephson currents $J(\varphi_{0})=\frac{2\pi}{\Phi_{0}}\partial_{\varphi_{0}}E(\varphi_{0})$ corresponding to Panel a). Panel c): The phase-dispersion of the CdGM levels in a system with a single flux quantum. Panel d): Separate contributions of individual CdGM states to the current in the junction with a single flux quantum. The simulations were performed with the same parameters as in Fig. \ref{fig: 1impurity}.}
\label{fig: 1impurity_currents}
\end{figure}

\section{Current operator}
\label{App:CurrentOperator}

We consider the projection of the current density operator in the normal part of the system ($x\in[0, W]$) near $y\approx y_{k}$
\begin{align}
    \frac{ve}{2}\Psi^{\dagger}(x, y)\sigma_{x}\Psi(x, y)\approx \frac{ve}{2}\sum_{\sigma=\pm}\sigma|\psi_{\sigma, k}(x, y)|^{2}|\alpha_{\sigma, k}(y)|^{2}=\frac{ve}{2}\frac{\kappa_{k}(y)}{1+\kappa_{k}(y)W(y)}\sum_{\sigma=\pm}\sigma|\alpha_{\sigma, k}(y)|^{2}.
\end{align}
Further one uses that $\frac{\partial \varepsilon_{k}(y)}{\partial y}=\frac{\pi}{\ell_{B}}\frac{v\kappa_{k}(y)}{1+\kappa_{k}(y)W(y)}$ to recover the result stated in the main text. 

\section{Perturbative expansion}
\label{App:PerturbativeExpansion}

We treat the last term of (\ref{eq:hkeff}) perturbatively, i.e., we split $h_{k}^{\text{eff}} = H_0 + V$, where $H_0 \equiv -i \bar v_k \rho_y\partial_y  +\alpha_k\cdot (y-y_k)\rho_z$ and $V\equiv \frac{\gamma_k}{2}\left\{-i\partial_y,(y-y_k)\right\}\rho_{y}$.

A standard procedure
\begin{align}
y-y_k = \sqrt{\frac{\bar v_k}{2\alpha_k}}(a^{\dag}+a)\ ;\ 
-i\partial_y = i\sqrt{\frac{\alpha_k}{2\bar v_k}}(a^{\dag}-a)
\end{align}
leads to
\begin{align}
H_{0} = \frac{\omega_{Bk}}{2}\left[(a^{\dag}+a)\rho_{z} + i(a^{\dag}-a)\rho_{y}\right]\ .
\end{align}
Here $\omega_{Bk}\equiv \sqrt{2\alpha_k \bar v_k}$. For the perturbation we obtain
\begin{align}
V= \frac{i\gamma_k}{2}\,\rho_{y}\left({a^{\dag}}^{2}- a^{2}\right)\ .
\end{align}

The eigenstates and the eigenvalues of $H_{0}$ are
\begin{align}
\ket{\Psi_{0}} =i \ket{0} \ket{\downarrow_{x}}\quad{\rm with}\quad E_{0}=0\ ,
\end{align}  
(the factor $i$ is needed to make this state invariant under charge conjugation $C=\rho_x K$)
and
\begin{align}
&\ket{\Psi_{n}} = \frac{1}{\sqrt{2}} \big(\ket{n-1}\ket{\uparrow_{x}} + \ket{n}\ket{\downarrow_{x}}\big)\ , \\
&\ket{\Psi_{-n}} = \frac{1}{\sqrt{2}} \big(\ket{n-1}\ket{\uparrow_{x}} - \ket{n}\ket{\downarrow_{x}}\big)\ ,
\end{align}  
with $E_{n,k} = - E_{-n,k} = \omega_{Bk} \sqrt{n}$. Here $n>0$. 
The states $\ket{n} \equiv \phi_n(y-y_k)$ are the standard eigenstates of a harmonic oscillator. As required $\ket{\Psi_{-n}} = C\ket{\Psi_{n}}$.

For the current profiles of the unperturbed states we obtain $J_0(y)=0$ and
\begin{align}
J_n(y) = \bra{\Psi_{n}} j_{k}(y) \ket{\Psi_{n}} =\mathcal{J} \phi_{n-1}(y-y_k)\phi_n(y-y_k)\ ,
\end{align}
where $\mathcal{J}\equiv \ell_B\alpha_k/\Phi_0$. All these profiles integrate to zero, as expected. 

We perform the standard first order perturbation expansion in $\gamma_k$ and obtain the corrections $\delta J_n(y)$.
The first two corrections read
\begin{align}
\delta J_0(y) &\equiv 0,\\
\delta J_1(y) &=  -\frac{\mathcal{J}\gamma_k}{6\omega_{Bk}} \left(6\phi_1^2-\sqrt{6}\phi_1\phi_3 + 3\sqrt{2}\phi_0\phi_2-2\sqrt{6}\phi_0\phi_4 \right)\ .
\end{align}
The arguments of the $\phi_n$ function on the RHS are all $y-y_k$.

\end{widetext}

\end{document}